\documentclass[12pt]{article}
\usepackage{amssymb,amsmath}
\usepackage[noblocks]{authblk}
\usepackage[top=0.75in, bottom=0.75in, left=0.75in, right=0.75in, dvips]{geometry}
\usepackage{caption}
\begin{document}
\textwidth 10.0in 
\textheight 9.0in 
\topmargin -0.60in
\title{Path Integral Quantization of the First Order Einstein-Hilbert Action from its Canonical Structure}
\author[1,3]{Farrukh Chishtie} \author[1,2]{D.G.C. McKeon}
\affil[1] {Department of Applied Mathematics, The
University of Western Ontario, London, ON N6A 5B7, Canada}
\affil[2] {Department of Mathematics and
Computer Science, Algoma University, Sault St.Marie, ON P6A
2G4, Canada}
\affil[3] {Department of Space Science, Institute of Space Technology, Islamabad 44000, Pakistan}
\maketitle

\maketitle

\begin{abstract}
We consider the form of the path integral that follows from canonical quantization and apply it to the first order form of the Einstein-Hilbert action in $d > 2$ dimensions.  We show that this is inequivalent to what is obtained from applying the Faddeev-Popov (FP) procedure directly. Due to the presence of tertiary first class constraints, the measure of the path integral is found to have a substantially different structure from what arises in the FP approach. In addition, the presence of second class constraints leads to non-trivial ghosts, which cannot be absorbed into the normalization of the path integral. The measure of the path integral lacks manifest covariance.
\end{abstract}
\noindent
PACS Keywords: Gravity, higher-dimensional, 04.50.-h, Quantum gravity, 04.60.-m, Quantum field theory, 03.70.+k, 11.10.-z

\section{Introduction}

When quantizing the Yang-Mills model of non-Abelian vector gauge Bosons, the Faddeev-Popov (FP) procedure [1] provides a way of maintaining manifest Lorentz covariance.  However, this approach to quantization is not obviously equivalent to canonical quantizations.  It is only by showing that the path integral which follows from canonical quantization is in fact the same as the one that follows from the FP approach can this equivalence be established [2, 3].

When confronted with the problem of using the path integral to quantize the first order Einstein-Hilbert (1EH) action (often credited to Palatini [4], but in fact due to Einstein [5]), it is tempting to simply use the FP procedure.  The demonstration in ref. [6] that the two path integrals are equivalent relies on a canonical analysis of the 1EH action that does not involve tertiary constraints--though such constraints are necessary if one is to generate the diffeomorphism transformations that involve second derivatives.  An analysis of the canonical structure of the 1EH action that reveals the presence of these tertiary first class constraints (as well as second class constraints) appears in refs. [7, 8].

The 1EH is of particular interest, as the interaction term in the action is just cubic in the independent fields (the metric $g_{\mu\nu}$ and the affine connection $\Gamma_{\mu\nu}^\lambda$).  In contrast, when using the second order Einstein-Hilbert (2EH) action, the interactions are an infinite series in $g_{\mu\nu}$ [9, 10]; this considerably complicates the Feynman rules [6, 11]. (A simplification of the Feynman rules also occurs when using the first order form of the action for Yang-Mills theory [12].)

We shall employ the form of the path integral that follows from the canonical quantization procedure [13], but treat the constraints that arise in the 1EH action in a non-standard way.  The usual way of handling first class constraints appears in ref. [2]; this approach is adapted to deal with second class constraints in refs. [14, 15].

In this paper, the ``second class'' constraints must first be treated in an unusual way, as the ``first class'' constraints are only determined by considering a ``reduced'' action derived by using these second class constraints to eliminate fields from the 1EH action.  The 1EH action can be recovered from this reduced action within the path integral formalism provided a functional determinant contributes to the measure of the path integral which is distinct from the usual function determinant arising from second class constraints [14, 15].

For the first class constraints, we employ a technique that is adapted to the path integral in phase space from the FP approach to the path integral in configuration space for models possessing a gauge invariance [16].  That is, we integrate over ``gauge orbits'' that are generated by the first class constraints in the theory.  This approach allows us to employ a gauge fixing that is covariant.  The original approach in ref. [2] to quantize models containing first class constraints through use of the path integral requires an extension [17] in order to deal with covariant gauge fixing when working in phase space.

In the following section we review the canonical structure of this action and in the third section, we describe how to quantize the 1EH action using the path integral.  In Appendix A, we give a general discussion of the path integral for constrained systems.

\section{The Canonical Structure of the First Order Einstein-Hilbert Action}

The canonical structure of the 1EH action has been analyzed in refs. [7, 8] following the Dirac constraint procedure [18].  In this section, we review the approach detailed in ref. [7] for obtaining this canonical structure.

The Lagrangian in $d$ dimensions is 
\[ \mathcal{L}_d = h^{\mu\nu} \left( G_{\mu\nu ,\lambda}^\lambda + \frac{1}{d-1} G_{\lambda\mu}^\lambda G_{\sigma\nu}^\sigma - G_{\sigma\mu}^\lambda G_{\lambda\nu}^\sigma \right)\eqno(1) \]
where $h^{\mu\nu}$ and $G_{\mu\nu}^\lambda$ are related to the metric $g_{\mu\nu}$ and the affine connection $\Gamma_{\mu\nu}^\lambda$ through 
\[ h^{\mu\nu} = \sqrt{-g} \,g^{\mu\nu} \qquad (\det h^{\mu\nu} = - (-g)^{-1+d/2})\eqno(2) \]
\[\hspace{-2.2cm} G_{\mu\nu}^\lambda = \Gamma_{\mu\nu}^\lambda  - \frac{1}{2} \left( \delta_\mu^\lambda \Gamma_{\sigma\nu}^\sigma + \delta_\nu^\lambda \Gamma_{\sigma\mu}^\sigma\right).\eqno(3)\]
The infinitesimal form of the diffeomorphism invariance present in $\mathcal{L}_d$ is
\[ \delta h^{\mu\nu} = h^{\mu\lambda} \partial_\lambda \theta^\nu + h^{\nu\lambda} \partial_\lambda \theta^\mu -\partial_\lambda (h^{\mu\nu} \theta^\lambda)\eqno(4) \]
\[ \delta G_{\mu\nu}^\lambda = -\partial_{\mu\nu}^2 \theta^\lambda + \frac{1}{2} \left(\delta_\mu^\lambda \partial_\nu + \delta_\nu^\lambda \partial_\mu\right)\partial \cdot \theta - \theta \cdot \partial G_{\mu\nu}^\lambda \eqno(5) \]
\[ \hspace{1cm}+ G_{\mu\nu}^\rho \partial_\rho \theta^\lambda - \left(G_{\mu\rho}^\lambda \partial_\nu + G_{\nu\rho}^\lambda \partial_\mu\right) \theta^\rho.\nonumber \]

The Lagrangian is now expressed in terms of variables that simplifies the canonical analysis.  We initially take 
\[ h=h^{00},\quad h^i = h^{0i},\quad  H^{ij} = hh^{ij} - h^i h^j \eqno(6,7,8) \]
and
\[ \pi = - G_{00}^0, \quad \pi_i = -2 G_{0i}^0, \quad \pi_{ij} = -G_{ij}^0 \eqno(9,10,11) \]
\[ \xi^i = - G_{00}^i, \quad \xi_j^i = -2 G_{0j}^i, \quad \xi_{jk}^i = -G_{jk}^i .\eqno(12,13,14) \]
We then set
\[\hspace{-3.35cm}\pi_{ij} = h\Pi_{ij} \eqno(15) \]
\[\hspace{-1.8cm} \pi_i = \Pi_i - 2\Pi_{ij}h^j\eqno(16) \]
\[ \pi = \Pi + \frac{\Pi_{ij}}{h} (H^{ij} + h^i h^j) \eqno(17) \]
\[\hspace{-1.5cm} \overline{\xi}^i = \xi^i - \xi_{jk}^i h^{jk}/h \eqno(18) \]
\[ \xi_j^i = \overline{\zeta}_j^i - \frac{2}{h} \left( \xi_{jk}^i - \frac{1}{d-1} \delta_j^i \xi_{\ell k}^\ell\right) h^k\eqno(19) \]
\[\hspace{3cm}+ \delta_j^i \left( \overline{t} - \frac{1}{d-1} \frac{2}{h} \xi_{\ell k}^\ell h^k \right)\;\;(\overline{\zeta}_i^i = 0 )\nonumber \]
which leads to 
\[ \mathcal{L}_d = \Pi h_{,0} + \Pi_i h_{,0}^i + \Pi_{ij} H_{,0}^{ij} + \overline{\xi}^i \chi_i + \overline{t} \chi\eqno(20) \]
\[\hspace{3cm} + \frac{2-d}{d-1} \bigg[ h\Pi^2 + h^i \Pi\Pi_i + \frac{1}{4h} (H^{ij} + h^ih^j) \Pi_i \Pi_j \nonumber \]
\[\hspace{2cm} + \frac{1}{h} (H^{ij} H^{k\ell} + H^{ik} h^jh^\ell ) \Pi_{ij} \Pi_{k\ell} \nonumber \]
\[\hspace{3.4cm} + \frac{1}{h} (h^i H^{k\ell} - H^{ik} h^\ell ) \Pi_{i} \Pi_{k\ell} + 2H^{ij} \Pi\,\Pi_{ij}\bigg] \nonumber \]
\[\hspace{3.4cm} + \overline{\zeta}_j^i \left( h_{,i}^j - \frac{1}{2} h^j \Pi_i - H^{jk} \Pi_{ik} \right) - \frac{h}{4} \overline{\zeta}_j^i \overline{\zeta}_i^j \nonumber \]
\[\hspace{4cm} + \xi_{jk}^i \bigg[ \frac{1}{h} H_{,i}^{jk} - \frac{1}{h} H^{jk} \Pi_i + \frac{1}{2(d-1)h} (\delta_i^j H^{k\ell} + \delta_i^k H^{j\ell})\nonumber \]
\[\hspace{4cm}(\Pi_\ell - 2h^m \Pi_{\ell m} ) + \frac{1}{h} (h^j H^{kp} + h^k H^{jp} ) \Pi_{ip} \bigg] \]
\[\hspace{3cm} + \frac{1}{h} H^{ij} \left( \frac{1}{d-1} \xi_{ki}^k \xi_{\ell j}^\ell - \xi_{\ell i}^k \xi_{kj}^\ell \right)\nonumber \]
where 
\[ \chi_i = h_{,i} - h\Pi_{i} \eqno(21) \]
and
\[ \chi = h_{,i}^i + h\Pi . \eqno(22) \]
From this form of $\mathcal{L}_d$, we see that the canonical momenta conjugate to $\Pi$, $\Pi_i$ and $\Pi_{ij}$ all vanish while the canonical momenta conjugate to $h$, $h^i$ and $H^{ij}$ are $\Pi$, $\Pi_i$ and $\Pi_{ij}$ respectively.  These momenta constitute $d(d-1)$ primary second class constraints on these variables. The momenta conjugate to $\overline{\xi}^i$, $\overline{t}$, $\overline{\zeta}_j^i$ and $\xi_{jk}^i$ also vanish; these are the primary constraints
\[ I\!\!P_i = I\!\!P = \overline{I\!\!P}_i^j = {I\!\!P}_i^{jk} = 0 \eqno(23a-d)\]
respectively.  (We note that since $\overline{\zeta}_i^i = 0$, we must use the Poisson Bracket (PB)
\[ \left\lbrace \overline{\zeta}_j^i(\vec{r},t),\;\; \overline{I\!\!P}_k^\ell (\vec{r}^{\,\prime} ,t)\right\rbrace = \left(\delta_k^i \delta_j^\ell - \frac{1}{d-1} \delta_j^i \delta_k^\ell \right) \delta^{d-1} (\vec{r} - \vec{r}^{\,\prime}). )\eqno(24) \]

As the form of $\mathcal{L}_d$ is $p_i \dot{q}^i - H_c$, one can immediately read off the canonical Hamiltonian $\mathcal{H}_c$ from eq. (20).  We find that the constraints $I\!\!P_i = I\!\!P = 0$ lead to the secondary constraints
\[ \chi_i = \chi = 0 \eqno(25a,b) \]
while $\overline{I\!\!P}_i^j = I\!\!P^{jk}_{\,\,i} = 0$ lead to 
\[ \overline{\zeta}_j^i = \frac{2}{h}\left( \lambda_j^i - \frac{1}{d-1} \delta_j^i \lambda_k^k \right)\eqno(26) \]
\[ \xi_{jk}^i = -\frac{1}{2}(M^{-1})_{jk\;\;mn}^{\,i\;\;\;\;\;\;\ell} \sigma_\ell^{mn}\eqno(27)\]
where
\[ \lambda_i^j = h_{,i}^j - \frac{1}{2} h^j \Pi_i - H^{jk}\Pi_{ik}\eqno(28)\]
\[ \sigma_i^{jk} = \frac{1}{h} H_{,i}^{jk} - \frac{1}{h} H^{jk} \Pi_i + \frac{1}{2(d-1)h} \left(\delta_i^j H^{k\ell} + \delta_i^k H^{j\ell}\right) (\Pi_\ell - 2h^m\Pi_{\ell m})\nonumber \]
\[ + \frac{1}{h} \left( h^j H^{kp} + h^k H^{jp}\right)\Pi_{ip} \eqno(29) \]
and
\[(M^{-1})_{yz\;\;\ell m}^{\,x\;\;\;\;\;\;k} = - \frac{h}{2}\bigg[ \left( H_{\ell y} \delta_z^k \delta_m^x + H_{my}\delta_z^k\delta_\ell^x + H_{\ell z} \delta_y^k \delta_m^x + H_{mz} \delta_y^k \delta_\ell^x\right) \eqno(30) \]
\[+ \frac{2}{d-2} \left( H^{k x} H_{\ell m} H_{yz}\right) - H^{kx} 
 \left( H_{\ell z} H_{my} +  H_{mz} H_{\ell y}\right)\bigg] .\nonumber \]
Eqs. (23c,d) and (26,27) obviously constitute a set of $d(d^2 - 3)$ second class constraints.  The secondary constraints of eqs. (25a,b) do not have vanishing PB with those of eqs. (26,27).  However, this does not mean that all these constraints are second class.  We take the constraints of eqs. (23c,d) and (26,27) to be second class and use them to eliminate the fields $\overline{\zeta}_j^i$ and $\xi_{jk}^i$ from $\mathcal{H}_c$.  (The Dirac Brackets that replace the Poisson Brackets are identical to the Poisson Brackets for the variables that are left after this replacement [7].)  The resulting expression for $\mathcal{H}_c$ is 
\[ \mathcal{H}_c = \frac{1}{h} (\tau + h^i \tau_i) + \frac{3}{2(d-2)}\,\frac{1}{h^2} H_{,i}^{k\ell} H_{k\ell} H^{ij} \chi_j \eqno(31) \]
\[\hspace{3.2cm} -\frac{3d-5}{4(d-2)} \,\frac{1}{h^3} H^{ij} \chi_i\chi_j - \frac{3}{h^2} H^{ij} \chi_i \Pi_j - \left( \frac{1}{h^2} H^{ij} \chi_i\right)_{,j}\nonumber \]
\[ \hspace{3.4cm}+ \frac{2}{h^2} H_{,i}^{ij} \chi_j + \frac{h^i}{h} \chi_{,i} - \frac{1}{h^2} h^i h^j_{,i}\chi_j - 2 \left(\frac{d-2}{d-1}\right) \frac{1}{h} H^{mn} \Pi_{mn}\chi \nonumber \]
\[ \hspace{3cm}- \frac{1}{h} h^i \Pi\chi_i - \frac{H^{mn}\Pi_{mn}}{h^2} h^i\chi_i - \frac{2}{h^2} h^k H^{ij} \Pi_{ik}\chi_j\nonumber \]
\[\hspace{3.2cm} + \frac{d-3}{d-1} \Pi\chi + \frac{1}{d-1}\,\frac{\chi^2}{h} - \frac{1}{d-1}\,\frac{1}{h} h^i \Pi_i \chi - \overline{\xi}^i \chi_i - \overline{t}\chi\nonumber \]
where
\[\hspace{-4cm}\tau = H^{ij}_{,ij} - \frac{1}{2} H_{,n}^{mi}H_{ij} H_{,m}^{nj} - \frac{1}{4} H^{ij} H_{mn,i} H_{,j}^{mn} \eqno(32) \]
\[\hspace{1.4cm} - \frac{1}{4(d-1)} H^{ij} H_{k\ell} H_{,i}^{k\ell} H_{mn} H_{,j}^{mn} + H^{ij} H^{k\ell} \left( \Pi_{ij} \Pi_{k\ell} - \Pi_{ik}\Pi_{j\ell}\right)\nonumber \]
and
\[ \tau_i = -2(H^{mn} \Pi_{mi})_{,n} + H^{mn} \Pi_{mn,i} + (H^{mn} \Pi_{mn})_{,i}.\eqno(33) \]
(Eqs. (32,33) correct misprints in eqs. (46, 47) of ref. [7].)

With the Hamiltonian of eq. (31), we find that $(I\!\!P_i, I\!\!P)$, $(\chi_i, \chi)$ and $(\tau_i, \tau)$ are all first class constraints of the first, second and third generation respectively; the only non vanishing Poisson Brackets for these constraints are 
\[ \left\lbrace \chi_i, \chi \right\rbrace = \chi_i \eqno(34)\]
\[ \left\lbrace \tau_i(\vec{x}), \tau_j(\vec{y}) \right\rbrace = \left( -\tau_i (\vec{y}) \partial_j^x + \tau_j(\vec{x})\partial_i^y\right) \delta^{d-1}(\vec{x} - \vec{y})\eqno(35) \]
\[ \left\lbrace \tau(\vec{x}), \tau(\vec{y}) \right\rbrace = \left( \partial_i^x H^{ij}(\vec{y})\tau_j (\vec{y}) -  \partial_j^y  H^{ij}(\vec{x})\tau_i(\vec{x})\right)\delta^{d-1}(\vec{x}- \vec{y})\eqno(36) \]
and
\[ \left\lbrace \tau_i(\vec{x}), \tau(\vec{y}) \right\rbrace = \left( - \partial_i^x \tau(\vec{y}) +\partial_i^y \tau(\vec{x})\right) \delta^{d-1}(\vec{x} - \vec{y}).\eqno(37) \]

We now note that in phase space there are $d (d-3)$ degrees of freedom. (This equals four when $d=4$, they are the two polarizations of the graviton and their conjugate momenta. It is zero when $d=3$ which is expected as then the 2EH action is topological). To see this note that initially there are $d(d+1)^2$ fields in phase space ($h^{\mu \nu}$, ${G^{\lambda}}_{\mu \nu}$ and their conjugate momenta). There are $d(d+1)$ primary second class constraints associated with momenta conjugate to ($h$, $h^i$, $H^{ij}$). There are also $d(d^2-3)$ second class constraints given by eqs. (23c, 23d, 26, 27), $3d$ first class constraints given by eqs. (23a, 23b, 25a, 25b, 32, 33) which in turn lead to $3d$ gauge conditions. With these $d(d+1)+d(d^2-3) + 3d+3d$ restrictions, these are but $d(d-3)$ independent degrees of freedom left in the initial $d(d+1)^2$ dimensional phase space.

From the first class constraints, the generator of the gauge transformations that leave the extended action in phase space invariant, either by using the approach of HTZ [20, 21] or C [22, 23].  In the HTZ approach, the form of the generator $G$ for a transformation $\delta\Phi = \left\lbrace \Phi , G \right\rbrace$ is
\[ G = \int d^{d-1} x[\lambda^{a_{i}} \phi_{a_{i}}] \eqno(38) \]
where $\phi_{a_{i}}$ is the set of first class constraints of the $i$$^{th}$ generation $( i = 1 \ldots N)$.  This generator satisfies equation [20]
\[ \int d^{d-1}x \left[ \frac{D\lambda^{a_{i}}}{Dt}\phi_{a_{i}} + \left\lbrace G, H_c + U^{a_{i}}\phi_{a_{i}}\right\rbrace - \delta U^{a_{i}}\phi_{a_{i}} \right] = 0 \eqno(39)\]
where $D/Dt$ denotes a derivative with respect to time exclusive of dependence through the dynamical fields. To obtain the gauge generator that leaves the classical action invariant, one sets $U^{a_{i}}=\delta U^{a_{i}}=0$ ($i=2,...,N$) in eq. (39) (see eq. (A.12)).  For the 1EH action we have 
\[ G = \int d^{d-1} x \left[ a I\!\!P + a^i I\!\!P_i + b\chi + b^i \chi_i + c\tau + c^i \tau_i\right].\eqno(40)\]
Taking $c$ and $c^i$ to be exclusively functions of $t$, then eq. (39) is satisfied at order $\tau$ and $\tau_i$ respectively provided
\[ c_{,0} + \frac{b}{h} + \frac{(c^ih)_{,i}}{h^2} - \frac{ch_{,i}^i}{h} + \frac{(ch)_{,i}h^i}{h^2} = 0 \eqno(41) \]
and
\[ c_{,0}^i + \frac{h^i}{h}b  + b^i -  \frac{H^{ij}}{h^2} (ch)_{,j} + c_{,j}^i  \frac{h^j}{h} - c^j\left(\frac{h^i}{h}\right)_{,j} = 0. \eqno(42) \]
These equations fix $b$ and $b^i$ in terms of $c$ and $c^i$.  In principle, the coefficients of $\chi$ and $\chi_i$ in eq. (39) determine $a$ and $a^i$ while the coefficients of $I\!\!P$ and $I\!\!P_i$ fix $\delta U^{a_{1}}$.  However, as $\mathcal{H}_c$ in eq. (31) is quadratic in $\chi$ and $\chi_i$, $a$ and $a^i$ are not fixed uniquely by this procedure.  However, to compute $\delta h$, $\delta h^i$ and $\delta H^{ij}$ it is adequate to have $b$ and $b^i$ in terms of $c$ and $c^i$.  We find that
\[ \delta h = \left\lbrace h, G\right\rbrace = \left\lbrace h, \int d^{d-1} x (b\chi) \right\rbrace = -h^2c_{,0} - (hc)_{,j} h^j + ch\,h_{,j}^j - (hc^j)_{,j} \eqno(43) \]
\[ \delta h^i = \left\lbrace h^i, G\right\rbrace = \left\lbrace h^i, \int d^{d-1} x (b^j\chi_j) \right\rbrace = hc_{,0}^i - h^ihc_{,0} + h^ic h_{,j}^j  \eqno(44) \]
\[ - \frac{h^i(hc^j)_{,j}}{h} - (ch)_{,j} h^{ij} + c_{,j}^i h^j - c^j h_{,j}^i + \frac{c^jh^ih_{,j}}{h}\nonumber \]
and 
\[\hspace{-6.8cm} \delta H^{ij} = \left\lbrace H^{ij}, G\right\rbrace = 
\left\lbrace H^{ij}, \int d^{d-1} x(c\tau + c^k\tau_k)\right\rbrace \eqno(45) \]
\[\hspace{3cm} = 2 (H^{ij} H^{k\ell} - H^{ik} H^{j\ell}) \Pi_{k\ell} c + (H^{ik} c_{,k}^j + H^{jk} c_{,k}^i) - (H^{ij} c^k)_{,k} - c_{,k}^k H^{ij}. \nonumber \]
(Eqs. (43, 45) correct mistakes in eqs. (85, 87) of ref. [7].)  Eqs. (43, 44) are consistent with eq. (4) provided $\theta = -hc$ and $\theta^i = c^i - h^ic$.  However, for eq. (45) to be consistent with eq. (4) with these relationships between $(\theta, \theta^i)$ and $(c, c^i)$ we must eliminate $\Pi_{k\ell}$ in eq. (45) using the equations of motion 
\[ \dot{H}_{ij} = \left\lbrace H_{ij}, \int d^{d-1} x\mathcal{H}_c \right\rbrace =\left\lbrace\int H_{ij}, \int d^{d-1} x \left(\frac{1}{h} \right) (\tau + h^i \tau_i)\right\rbrace \eqno(46) \]
\[  = \frac{2}{h} (H^{ij} H^{k\ell} - H^{ik} H^{j\ell} ) \Pi_{k\ell} \nonumber \]
\[ \hspace{3cm} + \left[ H^{ik} \left( \frac{h^j}{h}\right)_{,k} + 
H^{jk} \left( \frac{h^i}{h}\right)_{,k} \right] \nonumber \]
\[\hspace{4cm}- \left( H^{ij} \frac{h^k}{h}\right)_{,k} - \left( \frac{h^k}{h}\right)_{,k} H^{ij} \nonumber \]
and 
\[ \chi = \chi_i = 0.\eqno(47) \]

Upon applying these on-shell conditions, we find (just as in ref. [28]) that eq. (45) is consistent with eq. (4).  We have not pursued the question of how $\delta G_{\mu\nu}^\lambda = \left\lbrace G_{\mu\nu}^\lambda , G\right\rbrace$ is related to eq. (5).

The method $C$ [22, 23] can also be used to find the generator of the gauge transformation in phase space associated with the 1EH action, as was considered in ref. [8].  In this approach, 
\[ G = \epsilon G_0 + \dot{\epsilon} G_1 + \ddot{\epsilon} G_2 \eqno(48) \]
with $G_A$ determined by the chain
\[\hspace{-2.4cm} G_2 \approx \mathrm{primary} \eqno(49)\]
\[ G_1 + \left\lbrace G_2, H_T \right\rbrace \approx \mathrm{primary} \nonumber \]
\[ G_0 + \left\lbrace G_1, H_T \right\rbrace \approx \mathrm{primary} \nonumber \]
\[  \left\lbrace G_0, H_T \right\rbrace \approx \mathrm{primary}. \nonumber \]
One begins with either $G_2 = I\!\!P$ or $G_2 = I\!\!P_i$ and then in both cases, finds the resulting expressions for $G_1$ and $G_0$ from eq. (49).  This however, is not a unique process as $\mathcal{H}_c$ in eq. (31) is quadratic in $\chi$ and $\chi_i$.  Furthermore, as was pointed out in ref. [22], the gauge transformation for the 2EH action generated by this $G$ is equivalent to a diffeomorphism (eqs. (4,5)) only on shell.

At this point, we would like to mention certain features of our canonical analysis of the 1EH action that are distinctive. First of all, in ref. [29] a canonical analysis of the 1EH action is presented which does not have the tertiary constraints $\tau$ and $\tau_i$ of eqs. (32,33). This is because in this reference, all equations of motion that do not involve time derivatives are used to eliminate fields from the Lagrangian before embarking on a canonical analysis; two of these equation (namely the trace of eq. (A.3) and eq.(A.4)) correspond to the secondary first class constraints $\chi$ and $\chi_i$ of eqs. (25a,b) respectively while the other two equations (eq. (A.2) and the traceless part of eq. (A.3)) correspond to the secondary second class constraints of eqs. (27, 26) respectively. The use of secondary first class constraints to eliminate fields from the action prior to performing a canonical analysis precludes being able to uncover any tertiary constraints, which we have shown to be necessary if one is to have a generator capable of generating gauge transformations involving the second derivative of gauge functions (as in eq.(5)). In Appendix A of ref. [7] it is also shown how in the first order action for spin-2 fields the same thing happens; one must not use equations of motion to eliminate fields if one is to obtain the tertiary constraints that are needed to find a gauge generator that can give rise to the second derivative of gauge functions. In ref. [6] the same procedure is used as in ref. [29] for classifying the constraints of the 1EH action (eq. (4.18) of [6] corresponds to eqs. (A.2-4) of ref. [29]). Consequently the Hamiltonian form of the path integral in ref. [6] does not involve either second class or tertiary first class constraints.

We also note our way of distinguishing first and second class constraints differs in detail from that of Dirac, both for the 1EH action and the first order spin-2 action (as in ref. [7].) It is evident that eqs. (23 a-d) are a set of primary constraints; these lead to the secondary constraints of eqs. (25a-b, 26, 27). It is immediately obvious that eqs. (23c-d, 26, 27) are second class as the PB's of \{$\overline{\zeta}_j^i$, $\overline{I\!\!P}_l^k$\} and \{$ \xi_{jk}^i$, $I\!\!P^{lm}_{\,\,n}$\} are non-zero. If one were to follow Dirac’s procedure exactly, we now would also conclude that  $\chi$ and $\chi_i$ of eq. (25) would also be second class as \{$\Pi_{i}$, $\lambda_j^k$\}, \{$\Pi$, ${\sigma_i}^{jk}$\} are non-vanishing. This would mean that all secondary constraints would be second class and that these are neither first class constraints nor tertiary constraints, thereby eliminating the possibility of the Dirac procedure leading to any gauge generator for that the 1EH section (or the spin-2 action). To circumvent this obvious shortcoming of making exact use of the Dirac procedure we modify it slightly: first the second class constrains of eqs. (23c-d, 26, 27) are eliminated by use of the appropriate PBs, then the constraints of eqs. (23 a-b, 25) are first class and imply the tertiary constraints of eqs. (32, 33). In this way the generator of the gauge transformation given in eq. (40) can be found and its relation to the diffeomorphism transformation of eq. (4,5) is described above.

\section{The Path Integral in Phase Space for the First Order Einstein-Hilbert Action}

In this section we will show how the action of eq. (1) can be quantized using the path integral whose general form is given by eq. (A.21).  The action for which we have the gauge invariance in phase space given by eqs. (43-45) is associated with the Hamiltonian of eq. (31).  However, this Hamiltonian is found by eliminating the fields $\overline{\zeta}_j^i$ and 
$\xi_{jk}^i$ from the original Hamiltonian implied by eq. (20) using eqs. (26, 27).  As discussed after eq. (30), we can classify the constraints of eqs. (23a, 23b, 25a, 25b, 32, 33) as being first class only after $\overline{\zeta}_j^i$, $\xi_{jk}^i$ have been eliminated using eqs. (26, 27).  Consequently it is appropriate to use the path integral of eq. (A.9) to quantize our action only if ``$H_T$'' is identified with the total Hamiltonian associated with the Hamiltonian of eq. (31).

As noted at the end of the preceding section, we have deviated in detail in the way in which the constraints of the theory are subdivided into first or second class. As a result, the contribution to the measure of the path integral coming from these constraints given in eqs. (A.4) and (A.5) must be modified as will be now described. In particular, we first will consider the constraints of eqs. (26-27) which, according to our approach, are second class.

Eqs. (26, 27) arise from those parts of the Lagrangian of eq. (20) of the form 
\setcounter{equation}{49}
\begin{equation}
Q_A f_A - \frac{1}{2} Q_A M_{AB} Q_B;
\end{equation}
so that with
\begin{equation}
Q_A = M_{AB}^{-1} f_B
\end{equation}
these terms become contributions to the Hamiltonian of eq. (31) of the form
\begin{equation}
-\frac{1}{2} f_A M_{AB}^{-1} f_B .
\end{equation}
Upon using the equation
\begin{equation}
e^{\frac{1}{2}f_AM_{AB}^{-1}f_B} = \left( \frac{\det M_{AB}}{(2\pi)^n}\right)^{1/2} \int D^n Q_A e^{Q_A f_A -\frac{1}{2} Q_A M_{AB}Q_B}
\end{equation}
it is apparent that $\overline{\zeta}_j^i$ and $\overline{\xi}_{jk}^i$ can be restored into the action and that we can use eq. (20) provided the contributions of $(\det M_{AB})^{1/2}$ coming from eqs. (26, 27) are included in the measure of the path integral.  These determinants are 
\begin{equation}
D_1(h) = \det \left[h\left(\delta_i^\ell \delta_k^j - \frac{1}{d-1} \delta_i^j \delta^\ell_k\right)\right]
\end{equation}
(coming from $-\frac{h}{4} \overline{\zeta}_j^i \overline{\zeta}_i^j$ in eq. (20)) and 
\begin{align}
D_2(H^{ij}/h) &= \det \bigg[\frac{1}{h}\bigg(\frac{1}{d - 1} \left( \delta_i^j \delta_\ell^m H^{kn} + \delta_i^k \delta_\ell^m H^{jn} + \delta_i^j \delta_\ell^n H^{km} + \delta_i^k \delta_\ell^n H^{jm}\right)\\
&  - \left( \delta_\ell^j \delta_i^m H^{kn} + \delta_\ell^k \delta_i^m H^{jn} + \delta_\ell^j \delta_i^n H^{km} + \delta^k_\ell \delta_i^n H^{jm}\right)\bigg)\bigg]\nonumber
\end{align}
(coming from $\frac{H^{ij}}{h}\left(\frac{1}{d-1} \xi_{ki}^k\xi_{\ell j}^\ell - \xi_{\ell i}^k \xi_{kj}^\ell\right)$ in eq. (20)). The contribution of the functional determinants
of eqs. (54, 55), arising from the contributions of the secondary second class constraints of eqs. (26, 27), is distinct from what one might expect from eq. (A.5), as eq. (A.5) follows from directly applying the Dirac procedure plus the analysis of refs. [14, 15].

Having the gauge invariances of eqs. (43-45) requires introduction of a factor of the form of eq. (A.10) into the path integral to remove the divergence arising from the ``overcounting'' of gauge-equivalent field configurations.  The gauge fixing term 
\begin{equation}
\psi^\nu = \partial_\mu h^{\mu\nu}
\end{equation}
is commonly used; it becomes using eqs. (6-8)
\begin{equation}
\psi^0 = \dot{h} + \partial_i h^i
\end{equation}
\begin{equation}
\psi^i = \dot{h}^i + \left( \frac{H^{ij} + h^ih^j}{h} \right)_{,j} .
\end{equation}
The results of eqs. (43-45) show that
\begin{align}
\left\lbrace \psi^0, G\right\rbrace &= \left( -h^2 c_{,0} - (hc)_{,j} h^j + ch\,h_{,j}^j - (h c^j)_{,j}\right)_{,0}\\
&\hspace{1cm}+ \big( hc_{,0}^i - h^i hc_{,0} + h^i ch_{,j}^j - \frac{1}{h} h^i(hc^j)_{,j} - (ch)_{,j}h^{ij} \nonumber \\
&\hspace{3cm}+ c^i_{,j} h^j - c^j h_{,j}^i + \frac{1}{h} c^j h^i h_{,j}\big)_{,i}\nonumber
\end{align}
and 
\begin{align}
\left\lbrace \psi^j, G\right\rbrace &= \big( hc_{,0}^i - h^ihc_{,0} + h^i ch_{,j}^j - \frac{1}{h} h^i (hc^j)_{,j} - (ch)_{,j} h^{ij} \nonumber \\
&\hspace{2cm}+ c_{,j}^i h^j - c^j h_{,j}^i + \frac{1}{h} c^jh^ih_{,j}\big)_{,0} \\
&\hspace{1cm} + \frac{1}{h} \bigg( 2(H^{ij} H^{k\ell} - H^{ik} H^{j\ell})\Pi_{k\ell} c + (H^{ik} c_{,k}^j + H^{jk} c_{,k}^i) \nonumber \\
&\hspace{2cm}-(H^{ij} c^k)_{,k} - c_{,k}^k H^{ij}\bigg)_{,j}\nonumber  \\
&\hspace{1cm}+ \frac{h^i}{h} \big( hc_{,0}^j - h^j hc_{,0} + h^j ch_{,k}^k - \frac{1}{h} h^j (hc^k)_{,k} - (ch)_{,k} h^{jk} \nonumber \\
&\hspace{2cm}+ c_{,k}^j h^k - c^k h_{,k}^j + \frac{1}{h} c^k h^j h_{,k}\big)_{,j} \nonumber \\
&\hspace{1cm}+ \frac{h^j}{h} \big( hc_{,0}^i - h^i hc_{,0} + h^i ch_{,k}^k - \frac{1}{h} h^i (hc^k)_{,k} - (ch)_{,k} h^{ik} \nonumber \\
&\hspace{2cm}+ c_{,k}^i h^k - c^k h_{,k}^i + \frac{1}{h} c^k h^i h_{,k}\big)_{,j} \nonumber \\
&\hspace{1cm}- \frac{H^{ij} + h^i h^j}{h^2} \left( -h^2 c_{,0} - (hc)_{,k} h^k + ch\,h_{,k}^k - (hc^k)_{,k}\right)_{,j}. \nonumber
\end{align}
The functional integral over $\mu_{a_{N}}$ in eq. (A.10) (identifying $\lambda^{a_{N}}$ with $c$ and $c^k$) gives rise to the determinant $\Delta$, since
\begin{equation}
\int d^4x \delta(\mathbf{A} \vec{x} + \vec{b}) = (|\det \,\mathbf{A}|)^{-1}.
\end{equation}
As
\begin{equation}
\det\mathbf{A} = \int d^n \gamma d^n\overline{\gamma} \exp (\overline{\gamma}^T \mathbf{A} \gamma)
\end{equation}
using Grassmann fields $\overline{\gamma}$, $\gamma$, we see that the contribution of eq. (59) to $\Delta$ can be exponentiated to give
\begin{align}
\Delta^{(1)} &= \int D\overline{\gamma}^\lambda D\gamma^\lambda \exp i \int dx \bigg\{ -\overline{\gamma}^0_{,0} \left[ -h^2 \gamma_{,0}^0 - (h\gamma^0)_{,j} h^j + \gamma^0 h\,h_{,j}^j - (h\gamma^j)_{,j}\right]\\
&\hspace{2cm} -\overline{\gamma}^0_{,i} \left[ h\gamma^i_{,0} - h^i h\gamma^0_{,0} + h^i\gamma^0 h_{,j}^j - \frac{1}{h} h^i(h\gamma^j)_{,j} - (\gamma^0 h)_{,j} h^{ij} \right. \nonumber \\
&\left.\hspace{3.4cm} -\gamma_{,j}^i h^j - \gamma^j h_{,j}^i + \frac{1}{h} \gamma^j h^i h_{,j}\right]\bigg\} \nonumber
\end{align}
with $\Delta^{(2)}$, coming from eq. (60), being exponentiated in exactly the same manner; we have $\Delta = \Delta^{(1)}\Delta^{(2)}$.
All together then, the path integral of eq. (A.14) becomes
\begin{align}
<\mathrm{out}|\mathrm{in}>& = \int Dh\,Dh^i\,DH^{ij}\, D\Pi\,D\Pi_i\,D\Pi_{ij}\, D\overline{\xi}^i\, D\overline{t}\, D\overline{\zeta}_j^i D\xi_{jk}^i\nonumber \\
&\hspace{1.3cm} \left( D_1(h)D_2(H^{ij})\right)^{1/2}\quad \Delta \quad \exp i\int d^dx\left( \mathcal{L}_d - \frac{1}{2\alpha} (\partial_\mu h^{\mu\nu})^2\right)
\end{align}
with $\mathcal{L}_d$ being given by eq. (20).

One can now revert back to the variables $h^{\mu\nu}$, $G_{\mu\nu}^\lambda$ in place of those that appear in eq. (64) as these two sets of variables are canonically equivalent.  This results in $\mathcal{L}_d$ being expressed in the form of eq. (1) rather than eq. (20).  The determinants $D_1$, $D_2$, $\Delta^{(1)}$ and $\Delta^{(2)}$ are similarly expressed in terms of $h^{\mu\nu}$, $G_{\mu\nu}^\lambda$.

We note that since $\mathcal{L}_d$ (both in eq. (20) and eq. (1)) is already in ``Hamiltonian form'' (viz of the form $p\dot{q}-H$) it is not necessary to employ the technique outlined in eqs. (A.14-22) to eliminate the path integral over momenta and obtain the configuration space form of the path integral.

\section{Discussion}

By having started with the canonical structure of the 1EH action [7, 8] , found by using the Dirac constraint formalism [18], we have been able to quantize this model using the quantum mechanical path integral.  The result is different from what is obtained using the FP approach [6].  This is because of the presence of the determinants $D_1$, $D_2$ and $\Delta$ in  eq. (64).  (Once these determinants are exponentiated using Grassmanian variables as in eq. (62), one obtains the new ghost contributions to the effective action.) Both $D_1$ and $D_2$ arise because of ``second class'' constraints in the theory, though they are second class only in the sense that the first class constraints are so classified only after these second class constraints have been used to remove the fields $\xi_{jk}^i$ and $\overline{\zeta}_j^i$ from the action; $D_1$ and $D_2$ arise when 
restoring these variables.  The determinant $\Delta$ arises through breaking of the gauge invariance present in the phase space form of the total action by virtue of the 3d first class constraints present in the theory. It is distinct from the usual FP determinant.

Although the Lagrangian $\mathcal{L}_d$ is manifestly covariant, neither these determinants nor the gauge choice $\partial_\mu h^{\mu\nu}$ have this  property.  The determinants $D_1$ and $D_2$ are not covariant, as their form arises upon a particular choice of a ``time'' coordinate; this is an inevitable feature of any canonical analysis. We have not been able to attribute any particular physical significance to the second class constraints that give rise to these two determinant; possibly this is due to their algebraic form arising from our choice of ``time'' coordinate. If one were to understand the underlying reason for the second class constraints then one might be better able to reconcile our canonical approach with manifest covariance. (The significance of the first class constraints is that they lead to a gauge generator $G$ whose effect is related to the diffeomorphism gauge transformation.) The determinant $\Delta$ is not covariant as the generator $G$ does not generate an infinitesimal diffeomorphism transformation in $h^{\mu\nu}$, but rather a transformation that becomes a diffeomorphism only if the gauge functions have a non-covariant field dependence and the equations of motion are satisfied (see the discussion following eq. (45)).  The gauge fixing term in eq. (64), $(\partial_\mu h^{\mu\nu})^2$, is not covariant, but the choice of gauge fixing $\psi$ in eq. (A.10) should be arbitrary.

It is not clear if Green's functions computed using eq. (A.10) would be manifestly covariant.  For Yang-Mills theory [16], quantization in phase space leads the same generating functional as is found by employing the FP approach, as in the Yang-Mills model there are no second class constraints and the first class constraints generate the usual Yang-Mills gauge transformations without requiring field dependent gauge functions.  The same is also true for the free spin two field and the 1EH action in two dimensions [16], though the same is not true for a model containing an anti-symmetric tensor field with a pseudo scalar mass term which is coupled to gauge field [19].

One generally uses the FP approach in conjunction with the configuration space form of the action in order to quantize a gauge theory in a manner consistent with manifest covariance.  Although this approach leads to a path integral that is consistent with the path integral that follows from canonical quantization for Yang-Mills theory [2, 3], it is not necessarily true for any gauge theory.  The argument for the equivalence of these two forms of the path integral for the 1EH action that appears in ref. [6] relies on a canonical analysis of this action that is inconsistent with the analysis of section two.  

Using the generating functional of eq. (64) to find a propagator associated with these terms in the action which are bilinear in the fields is not feasible, as is discussed in refs. [6, 16].  Only by expanding $h^{\mu\nu}$ about a flat background so that 
\begin{equation}
h^{\mu\nu} \rightarrow \eta^{\mu\nu} + h^{\mu\nu}
\end{equation}
can a suitable propagator be found with the gauge choice $\partial_\mu h^{\mu\nu} = 0$.  The expansion of eq. (65) can be done in the generating functional of eq. (64) directly.  However, it may be more appropriate to do this expansion in the action of eq. (1) and then performing the canonical analysis of the resulting action.  This latter course of action is non-trivial, but is currently being considered.

\section*{Acknowledgements}
S. Kuzmin and N. Kiriushcheva had useful comments, as did R.  Macleod.

\section*{Appendix}
\section*{The Path Integral for Constrained Systems}

If a system has degrees of freedom $(q_i(t), p^i(t))\;\;(i = 1 \ldots n)$ in phase space, and there are no constraints, then the equation of motion for a dynamical variable $A(q_i(t), p^i(t))$ is given by
\[ \frac{dA}{dt} = \left\lbrace A, H_c \right\rbrace \eqno(A.1) \]
where $H_c$ is the usual canonical Hamiltonian associated with the Lagrangian $L$ for the system 
\[ H_c (q_i, p^i) = p^i \dot{q}_i - L(q_i, \dot{q}_i). \eqno(A.2) \]
Upon canonical quantization, eq. (A.1) leads to the path integral for the generating functional [13]
\[ <\mathrm{out}|\mathrm{in}> = \int Dq_i\,Dp^i \exp i\int_{-\infty}^\infty dt(p^i \dot{q}_i - H_c) \eqno(A.3) \]
where $q_i(t) \rightarrow (q_{\mathrm{out}}, q_{\mathrm{in}})$ as $t \rightarrow \pm \infty$.

When first class constraints $\phi_a(q_i,p^i)$ are present, Faddeev [2] showed that the measure in eq. (A.3) is supplemented by a factor of 
\[ \det \left\lbrace \phi_a, \chi_b \right\rbrace \delta(\phi_a) \delta(\chi_a).\eqno(A.4) \]
where $\chi_a$ is the gauge condition associated with $\phi_a$.  This was done by performing a canonical transformation to explicitly eliminate those degrees of freedom that are not truly dynamical by virtue of the constraints that are present.  Senjanovic [15] extended these considerations, showing that if second class constraints $\theta_a$ are present, then the measure of eq. (A.3) receives a further supplement of
\[ \mathrm{\det}^{1/2} \left\lbrace \theta_a, \theta_b \right\rbrace \delta(\theta_a) \eqno(A.5) \]
in addition to that of eq. (A.4).  In many (but not all [19]) cases, this contribution to the measure coming from second class constraints is inconsequential.

We adopt a different approach [16] to the path integral for systems in which there are first class constraints $\phi_{a_{i}}$ where the subscript ``$i$'' denotes the generation ($i = 1$, primary; $i = 2$, secondary; $i = 3$, tertiary; $\ldots N$) of the constraint.  The primary constraints $\phi_{a_{i}}$ arise from the failure to be able to solve
\[ p^i = \frac{\partial L}{\partial \dot{q}_L} \eqno(A.6) \]
for $\dot{q}_i$ in terms of $q_i$ and $p^i$.  In this case, we define a ``total Hamiltonian'' $H_T$
\[ H_T = H_c + U^{a_{1}}\phi_{a_{1}} \eqno(A.7) \]
where $U^{a_{1}}$ are a set of Lagrange multipliers.  The Lagrangian equations of motion are now equivalent to 
\[ \frac{dA}{dt} = \left\lbrace A, H_T \right\rbrace \eqno(A.8a)\]
\[\phi_{a_{1}} = 0 \eqno(A.8b)\]
in place of eq. (A.1).  This leads to the path integral in phase space
\[ <\mathrm{out}|\mathrm{in}> = \int Dq_i\,Dp^i \,DU^{a_{1}} \exp i\int_{-\infty}^\infty dt(p^i \dot{q}_i - H_T). \eqno(A.9) \]
However, the presence of first class constraints implies the existence of a gauge invariance in the total action $S_T = \int (p^i \dot{q}_1 - H_T)$ and so the path integral in eq. (A.9) is ill defined.  We adapt the FP [1] approach, which they used in conjunction with path integrals in configuration space, to factor out the integration over degrees of freedom that are gauge artifacts in the phase space path integral of eq. (A.9).  This is done by first introducing a constant factor [16]
\[ 1 = \int D\lambda^{a_{N}} \delta(\psi + \left\lbrace \psi, G \right\rbrace - k) \Delta \eqno(A.10) \]
in eq. (A.9).  In eq. (A.10), $\psi(q_i,\dot{q}_i)$ fixes the gauge, $G$ is the generator of the gauge transformation that leaves $S_T$ invariant (found by using the HTZ method [20,21], as opposed to the $C$ method [22, 23]), $k$ is independent of $q_i$, and $\Delta$ is a functional determinant (analogous to the FP determinant [1]) introduced to ensure the equality in eq. (A.10).  The form of $G$ is 
\[ G = \sum_{i=1}^N \sum_{a_{i}} \lambda^{a_{i}} \phi_{a_{i}}(q_i, p^i);\eqno(A.11) \]
We assume that the system is such that the HTZ equation [20]
\[\frac{D\lambda^{a_{i}}}{Dt} \phi_{a_{i}} + \left\lbrace G, H_T \right\rbrace - \delta U^{a_{1}}\phi_{a_{1}} = 0 \eqno(A.12) \]
fixes $\lambda^{a_{1}} \ldots \lambda^{a_{N-1}}$ in terms of $\lambda^{a_{N}}$.

A further insertion of the constant
\[ \int Dk \exp \frac{-i}{2\alpha} \int_{-\infty}^\infty dt\, k^2 \eqno(A.13) \]
into eq. (A.9) followed by a gauge transformation generated by $(-G)$ leads to 
\[ <\mathrm{out}|\mathrm{in}> = \int D\lambda^{a_{N}} \int \,Dq_i \,Dp^i  
\,D\lambda^{a_{1}} \Delta \exp i\int_{-\infty}^\infty dt\left(p^i \dot{q}_i - H_T - \frac{1}{2\alpha} \psi^2\right)\eqno(A.14) \]
since the total action and $\Delta$ are gauge invariant.  The integral over  $\lambda^{a_{N}}$, which has been factored out, contains the divergence originally present in the path integral of eq. (A.9).

We now will follow the method outlined in refs. [24, 25, 26] to convert the integral in eq. (A.14) to an integral in configuration space.  If the rank of the $n \times n$ Hessian matrix
\[ A_r(q_i,\dot{q}_i) = \frac{\partial^2L(q_k,\dot{q}_k)}{\partial\dot{q}_i \partial\dot{q}_j}\eqno(A.15) \]
is $r$, then there are $n - r$ primary constraints $\phi_{a_{1}}(q_i,p^i)$.  
If we denote the first $r$ variable with a single prime 
($q^\prime_i$, $p^{\prime i}; i = 1 \ldots r$) and the remaining $n - r$ variables with a double prime $(q_i^{\prime \prime}, p_i^{\prime\prime}; i = r + 1 \ldots n)$, then from eq. (A.6) for $i = 1 \ldots r$
\[ \dot{q}^\prime_i = f_i(q^\prime_j, q_j^{\prime\prime},p^\prime_j, \dot{q}_j^{\prime\prime} ) \quad (i = 1 \ldots r)\eqno(A.16) \]
and 
\[ \phi_{a_{1}}(q_i, p^i) = p^{\prime\prime i} - g^i(q^\prime_j, q^{\prime\prime}_j, p^\prime_j) \; (i = r +1 \dots n) .\eqno(A.17) \]
By using eq. (A.2), (A.16) and (A.17), eq. (A.14) becomes 
\[\hspace{-2cm} <\mathrm{out}|\mathrm{in}> = \int Dq_i \,Dp^i \int Dv_i \delta \left(v_i - f_i (q_j, p_j^\prime, \dot{q}_j^{\prime\prime})\right)\Delta (q_i, v_i, \dot{q}_i^{\prime\prime})\nonumber \]
\[|A_r(q_i, v_i, \dot{q}_i^{\prime\prime})|\delta \left(\phi_{a_{1}}(q_i, p^i)\right) \exp i\int_{-\infty}^\infty dt \bigg[ p^{\prime i} ( \dot{q}^\prime_i - f_i) +  (p^{\prime\prime i} - g^i)\dot{q}_i^{\prime\prime} \eqno(A.18) \]
\[+ L(q_i, v_i, \dot{q}_i^{\prime\prime}) - \frac{1}{2\alpha} \psi^2 (q_i, v_i, \dot{q}_i^{\prime\prime})\bigg].\nonumber \]  

We now use the standard result
\[dx \delta (f(x)) = dx \sum_i \delta (x - a_i)/|f^\prime (a_i)|\qquad (f(a_i) = 0) \eqno(A.19) \]

to write 
\[ Dv_i \delta (v_i - f_i (q_j, p_j^\prime, \dot{q}_j^{\prime\prime})) = Dv_i |A_r (q_j, v_j, \dot{q}_j^{\prime\prime})| \eqno(A.20) \]
\[ \delta \left( p^{\prime i} - \frac{\partial L(q_j, v_j, \dot{q}_j^{\prime\prime}}{\partial v_i}\right). \nonumber \]

Eq. (A.20) serves to convert eq. (A.18) to 
\[ <\mathrm{out}|\mathrm{in}> = \int Dq_i \Lambda_r(q_i,\dot{q}_i) 
\exp i \int_{-\infty}^\infty dt L( q_i \dot{q}_i)\eqno(A.21) \]

where 

\[ \Lambda_r ( q_i ,\dot{q}_i) = \int Dv_i \exp \bigg[ - \frac{i}{2\alpha} \int_{-\infty}^\infty dt \psi^2 (q_i, v_i + \dot{q}_i^\prime, \dot{q}_i^{\prime\prime}) \bigg] \Delta \left(q_i, v_i + \dot{q}_i^\prime,  \dot{q}_i^{\prime\prime} \right)\nonumber \] 

\[| A_r( q_i, v_i + \dot{q}_i^\prime,  \dot{q}_i^{\prime\prime}) \delta \bigg[\phi_{a_{i}}\bigg(q_i, \frac{\partial L(q_j, v_j + \dot{q}_j^\prime,  \dot{q}_j^{\prime\prime})}{\partial v_i}\bigg),  
 \nonumber \]

\[ g^i \left(q_i,\frac{\partial L(q_j, v_j + \dot{q}_j^\prime,  \dot{q}_j^{\prime\prime})}{\partial v_i}\right)\bigg]       \eqno(A.22) \]

\[\exp \bigg[ i \int_{-\infty}^\infty dt \bigg(L (q_i, v_i + \dot{q}_i^\prime,  \dot{q}_i^{\prime\prime}) - L (q_i,  \dot{q}_i^\prime,  \dot{q}_i^{\prime\prime}) \nonumber \]
\[ - v_i \frac{\partial L(q_i, v_i + \dot{q}_i^\prime,  \dot{q}_i^{\prime\prime})}{\partial v_i}\bigg)\bigg] \nonumber \]
provided we make the shift $v_i \rightarrow v_i + \dot{q}^\prime$. 

We are considering how to handle systems (such as a scalar field on a curved background in $2D$ [27]) in which eq. (A.12) does not fix $\lambda^{a_{1}} \ldots \lambda^{a_{N-1}}$ in terms of $\lambda^{a_{N}}$.  We have also yet to examine how a gauge generator found using the approach of ref. [22] can be used in conjunction with eq. (A.10).

\end{document}